\documentclass[aps,prb,reprint,floatfix,amsmath]{revtex4-2}

\usepackage{graphicx}
\usepackage{latexsym}
\usepackage{xcolor}
\usepackage[utf8]{inputenc}
\usepackage[english]{babel}
\usepackage{lineno}

\usepackage{hyperref}

\DeclareMathOperator{\sign}{sgn}

\addto\captionsenglish{}

\usepackage{titlesec}

\makeatletter
\newcommand*{\smallrel}[2][.8]{%
  \mathrel{\mathpalette{\smallrel@{#1}}{#2}}%
}
\newcommand*{\smallrel@}[3]{%
  \sbox0{$#2\vcenter{}$}%
  \dimen@=\ht0 %
  \raise\dimen@\hbox{%
    \scalebox{#1}{%
      \raise-\dimen@\hbox{$#2#3\m@th$}%
    }%
  }%
}
\makeatother

\begin{document}
\setlength\linenumbersep{0.1cm}

\title{Feedback loop dependent charge density wave imaging by scanning tunneling spectroscopy}

\author{Alessandro Scarfato}
\thanks{To whom correspondence should be addressed: \href{mailto:alessandro.scarfato@unige.ch}{alessandro.scarfato@unige.ch}, \href{mailto:christoph.renner@unige.ch}{christoph.renner@unige.ch}}
\author{Árpád Pásztor}
\author{Lihuan Sun}
\author{Ivan Maggio-Aprile}
\author{Vincent Pasquier}
\author{Tejas Parasram Singar}
\author{Andreas Ørsted}
\author{Ishita Pushkarna}
\author{Marcello Spera}
\author{Enrico Giannini}
\author{Christoph Renner}

\thanks{To whom correspondence should be addressed: \href{mailto:alessandro.scarfato@unige.ch}{alessandro.scarfato@unige.ch}, \href{mailto:christoph.renner@unige.ch}{christoph.renner@unige.ch}}
\affiliation{Department of Quantum Matter Physics, University of Geneva, Quai Ernest-Ansermet 24, 1211 Geneva 4, Switzerland}

\date{\today}

\begin{abstract}
Scanning Tunneling Spectroscopy (STS) is a unique technique to probe the local density of states (LDOS) at the atomic scale by measuring the tunneling conductance between a sharp tip and a sample surface. However, the technique suffers of well-known limitations, the so-called set-point effect, which can potentially introduce artifacts in the measurements. 
We compare several STS imaging schemes applied to the  LDOS modulations of the charge density wave state on atomically flat surfaces, and demonstrate that only constant-height STS is capable of mapping the intrinsic LDOS.
In the constant-current STS, commonly used and easier-to-implement, the tip-sample distance variations imposed by the feedback loop result in set-point-dependent STS images and possibly mislead the identification of the CDW gap edges.
\end{abstract}

\maketitle

\section{Introduction}
The scanning tunneling microscope (STM) has established itself as a remarkable tool for characterizing the atomic and electronic structures of surfaces with unprecedented resolution. Scanning tunneling spectroscopy (STS) images and their Fourier transform, also known as Fourier transform STS (FT-STS)~\cite{Petersen2000}, give access to momentum resolution and band structure mapping through the analysis of quasiparticle interference (QPI) patterns~\cite{Avraham2018}. This information is extracted from the spatially resolved differential tunneling conductance, that under certain assumptions is proportional to the local density of states at the position of the tip apex.

The most common way to acquire STS images is to measure differential conductance spectra ($dI/dV(V)$) on a $(x,y)$ grid over an area of interest to obtain a four-dimensional $dI/dV(x,y,V)$ dataset ~\cite{hamers1986}. The second scheme consists of monitoring the output of a lock-in amplifier at the frequency of an applied AC modulation voltage ($V_\text{AC}$) while continuously scanning the tip along the surface at a fixed set point DC bias ($V_b$) ~\cite{Becker1985}. In both cases, the tip is scanned along the surface in constant-current mode (i.e. with an active feedback loop), meaning that the tip elevation is continuously adjusted to maintain a constant tunnel current. Since the tunnel conductance at a fixed location depends primarily on tip elevation and applied bias voltage, any spatial variation in the local density of states (LDOS) encountered while the tip scans the surface leads to a change in the tip elevation and, thus, in the tunnel barrier, which can introduce artifacts in the STS images.    

The impact of the feedback loop setting on STS imaging has been the subject of previous investigations, all suggesting that scanning the tip in constant height mode provides the most accurate mapping of the LDOS~\cite{krenner2013,Macdonald2016,Tresca2023}. MacDonald et al.~\cite{Macdonald2016} demonstrate that QPI on Ag(111) exhibits artifact features when scanning the tip in constant-current mode at a set-point close to or above the onset energy of its surface state. They conclude that to observe QPI features compatible with the parabolic surface state, either the set-point has to be well below the onset of the surface state band or the tip has to be scanned in constant-height mode. More recently, Tresca et al.~\cite{Tresca2023} have demonstrated that a controversial analysis of the charge density wave (CDW) structure in Pb/Si(111) is a direct consequence of the changing tip elevation during constant-current STS imaging.

Here, we discuss specific effects of open versus closed feedback loops on the CDW contrast inversion (CCI) observed in STS conductance maps, which has been demonstrated to inform about the width ($\Delta$) and position ($\epsilon$) of the CDW gap relative to the Fermi level ($E_F$) ~\cite{Spera2020}. We consider the commensurate $2a\times 2a$ CDW developing below 202~Kelvin in 1\textit{T}-TiSe$_2$. It is locked to the lattice and well described by a single CDW gap, avoiding the complications that may arise in materials presenting multiple CDW gaps~\cite{Cho2015, Pasztor2021}. Most importantly for the present study, $\Delta$ opens below $E_F$ in 1\textit{T}-TiSe$_2$, which, as we will show, results in a more complex CDW contrast as a function of energy, enabling a more complete discussion of the different STS acquisition modes. 

We find a strong dependence of the CDW contrast and its inversion on tunnelling bias and feedback loop setting. The observed effects range from additional CCIs not associated with any gap, to the complete absence of any CCI even though it is expected to occur across the CDW gap~\cite{Spera2020}. Structurally, the surface of 1\textit{T}-TiSe$_2$ is essentially flat. Hence, the very different bias dependencies of the STS contrast we measure are direct consequences of the feedback loop response to the LDOS modulated by the CDW reconstruction.  

\section{Experimental}
Single crystals of 1\textit{T}-TiSe$_2$ were grown via iodine vapour transport for 1 month at 590~°C to reduce the amount of self-intercalated Ti. They were cleaved in-situ in ultra-high vacuum at room temperature shortly before mounting them on the STM head. All STM and STS data were measured at 4.5~Kelvin using a SPECS Joule-Thomson STM with a base pressure better than $1\times10^{-10}$~mbar. The bias voltage was applied to the sample. Tips were mechanically cut from a PtIr wire and conditioned in-situ by Argon sputtering and scanning a clean Ag(111) surface. STS measurements were performed using a lock-in technique with $V_\text{AC}=3.54$~mV rms at 413.7~Hz. 

We compare three different STS imaging modes, mapping the local tunneling conductance $dI/dV$ as a function of energy, where $dI$ is the modulation amplitude of the tunneling current $I$ at the frequency of the applied  $V_\text{AC}$ measured with a lock-in amplifier. The first and most common STS mode is current-imaging tunneling spectroscopy (CITS) introduced by Hamers et al.~\cite{hamers1986}. In CITS, the tunneling conductance is measured over a set bias range with an open feedback loop at selected positions on a $(x,y)$ grid. Between sampling locations, the tip is moved in constant-current mode. The result of this mode, which we label constant-current CITS (CC-CITS), is a four-dimensional $dI/dV(x,y,V)$ data set.

Hamers et al.~\cite{hamers1986} developed CITS to overcome tip-sample distance dependencies of the measured LDOS, which they had already identified as a problem. However, CC-CITS is only a partial solution since there can still be different tip elevations between tunneling spectra measured at different locations. Here, we introduce constant-height CITS (CH-CITS), the second STS mode we examine, where the feedback loop is kept open during the entire STS map acquisition. CH-CITS also yields a four-dimensional $dI/dV(x,y,V)$ data set. The third STS mode we assess is constant-current conductance imaging (CC-CI), where the lock-in output is measured continuously while scanning the tip in constant-current mode. This results in a single $dI/dV(x,y,V_b)$ STS image, which has to be repeated for each desired bias voltage $V_b$. Note that CC-CI cannot be used to obtain very low or zero-bias conductance images because the tip would crash due to an insufficient tunnel current.

\section{\label{sec:model}Theoretical model}
To discuss our findings, we begin by outlining the one-dimensional (1D) model we use to reproduce the CCI observed in STM and STS images of a CDW. The starting point is the 1D Bardeen equation for the tunnel current at zero temperature~\cite{Chen1993}:

\begin{equation} 
    \label{eq:bardeen}
    I(x,z,V) \propto \int_{0}^{eV} \rho_t(E-eV) \rho_s(x,E) M(z,V,E) dE,
\end{equation}
where $\rho_t$ and $\rho_s$ are the tip and sample LDOS, respectively, $z$ is the tip-sample distance, $V$ is the tunnel bias voltage, and $M$ is the transmission factor. For a 1D trapezoidal barrier, $M$ can be written as:

\begin{equation}
    M(z,V,E) = e^{-\kappa(E,V) z},
\end{equation}
where $\kappa(E,V) ={\frac{2 \sqrt{2m}}{\hbar} \sqrt{\phi + \frac{eV}{2}-E}}$, and $\phi$ is the effective work function.

\begin{figure}
\includegraphics[width=0.95\columnwidth] {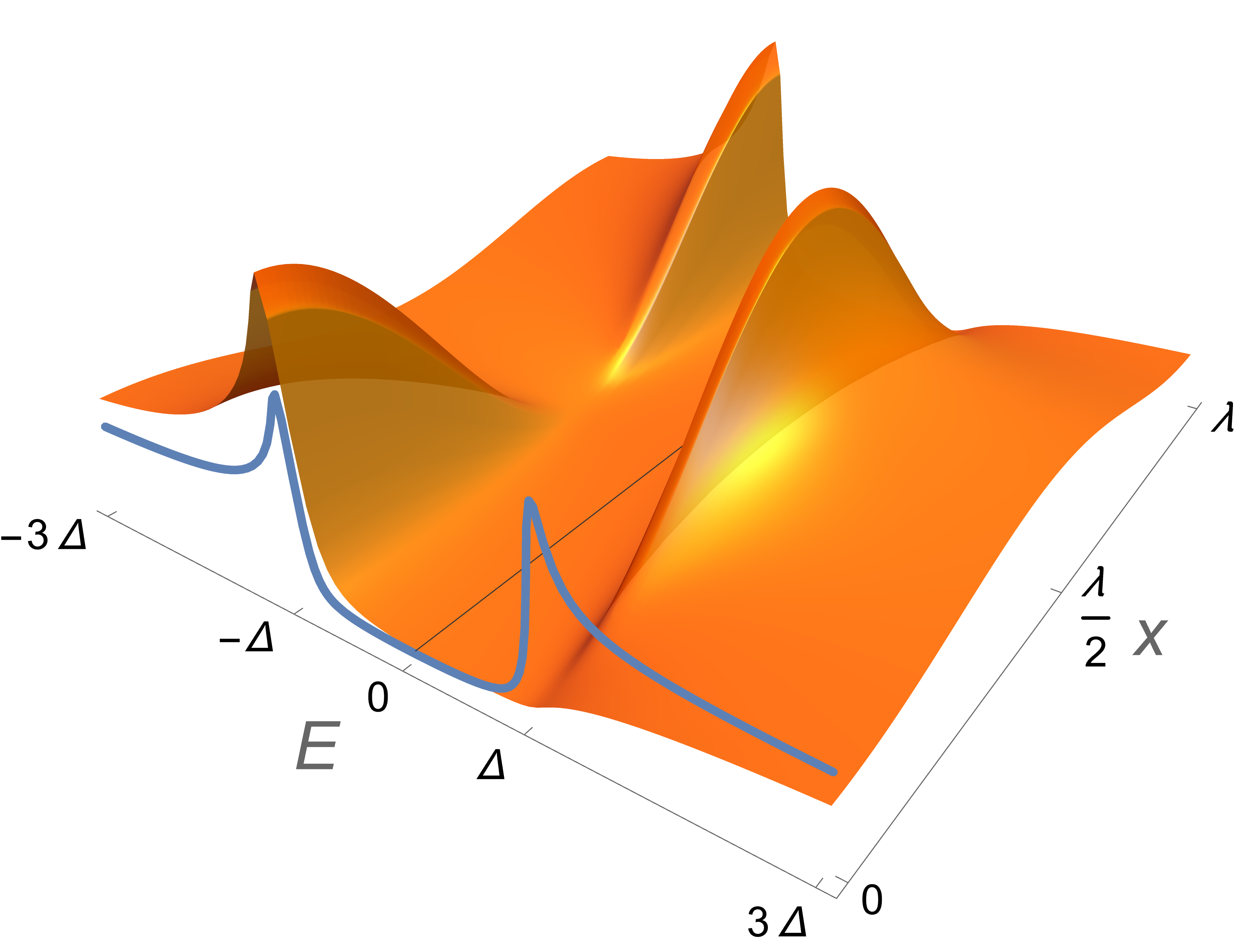}
\caption{Spatial dependence of the LDOS in the presence of a model CDW reconstruction in 1D, compared with a BCS superconducting gap (blue curve). The reconstructed CDW DOS is periodically modulated along the spatial dimension with period $\lambda$ and a relative shift of $\pi$ between the empty and filled states.
}
\label{fig:cdwdos}
\end{figure}

The LDOS corresponding to the CDW reconstruction (Fig.~\ref{fig:cdwdos}) is calculated using the mean field equation~\cite{Gruner1994,Dai2014}:

\begin{multline} 
    \label{eq:sampledos}
    \rho_s (x,E) = \rho_0 + \sign(E) \bigg[1 - \frac{\Delta}{E + i\mathit{\Gamma}} \cos{\frac{2 \pi x}{\lambda_\text{CDW}}} \bigg] \\ \frac{E + i\mathit{\Gamma}}{\sqrt{(E + i\mathit{\Gamma})^2 - \Delta^2}},
\end{multline}
where $E$ is the quasiparticle energy with respect to $E_F$, $\mathit{\Gamma}$ accounts for thermal and non-thermal broadening, $\lambda_\text{CDW}$ is the CDW wavelength, $\rho_0$ takes care of the fact that the CDW gap does not open for all momenta in the $(k_x,k_y)$ plane, and $\Delta$ is the width of the CDW gap, corresponding to the energy difference between the coherence peaks and the gap center, which we assume to be the same for all components in agreement with 1\textit{T}-TiSe$_2$ being a single gap CDW material.

Replacing $E$ by $E-\epsilon$ allows to account for a shift of $\Delta$ away from $E_F$ by an energy $\epsilon$. We assume a constant $\rho_t$, which can be taken out of the integral in Eq.~\eqref{eq:bardeen}. Since all the data discussed here were acquired at 4.5~Kelvin where $k_B T \ll \Delta$, we can use the zero temperature Bardeen equation for the tunnel current:

\begin{equation}
    \label{eq:bardeensimp}
    I(x,z,V) \propto \int_{0}^{eV} \rho_s(x,E) e^{-\kappa(E,V) z} dE,
\end{equation} 
with $\rho_s(x,E)$ and $\kappa(E,V)$ as defined above.

To analyse the set-point dependent STS images discussed below, we extend the 1D LDOS in Eq.\eqref{eq:sampledos} to an expression in 2D:

\begin{multline} 
    \label{eq:sampledos2d}
        \rho_s(x,y,E) = \rho_0 + \rho_{at} +\\ 
        \sign(E) \bigg[1 - \frac{\Delta}{E + i\mathit{\Gamma}} \sum_i \cos{(\mathbf{k}_i.\mathbf{r} + \phi_i)} \bigg] \\ 
        \frac{E + i\mathit{\Gamma}}{\sqrt{(E + i\mathit{\Gamma})^2 - \Delta^2}},
\end{multline}

where $\rho_{at}$ is a spatially modulated and energy-independent LDOS reproducing the atomic modulation and $\mathbf{k}_i$ are the independent wave vectors describing the 3Q CDW of 1\textit{T}-TiSe$_2$, with an angle
of 120° between each of them each with a phase $\phi_i$ \cite{McMillan1976}.

\begin{figure}
\includegraphics[width=1\columnwidth] {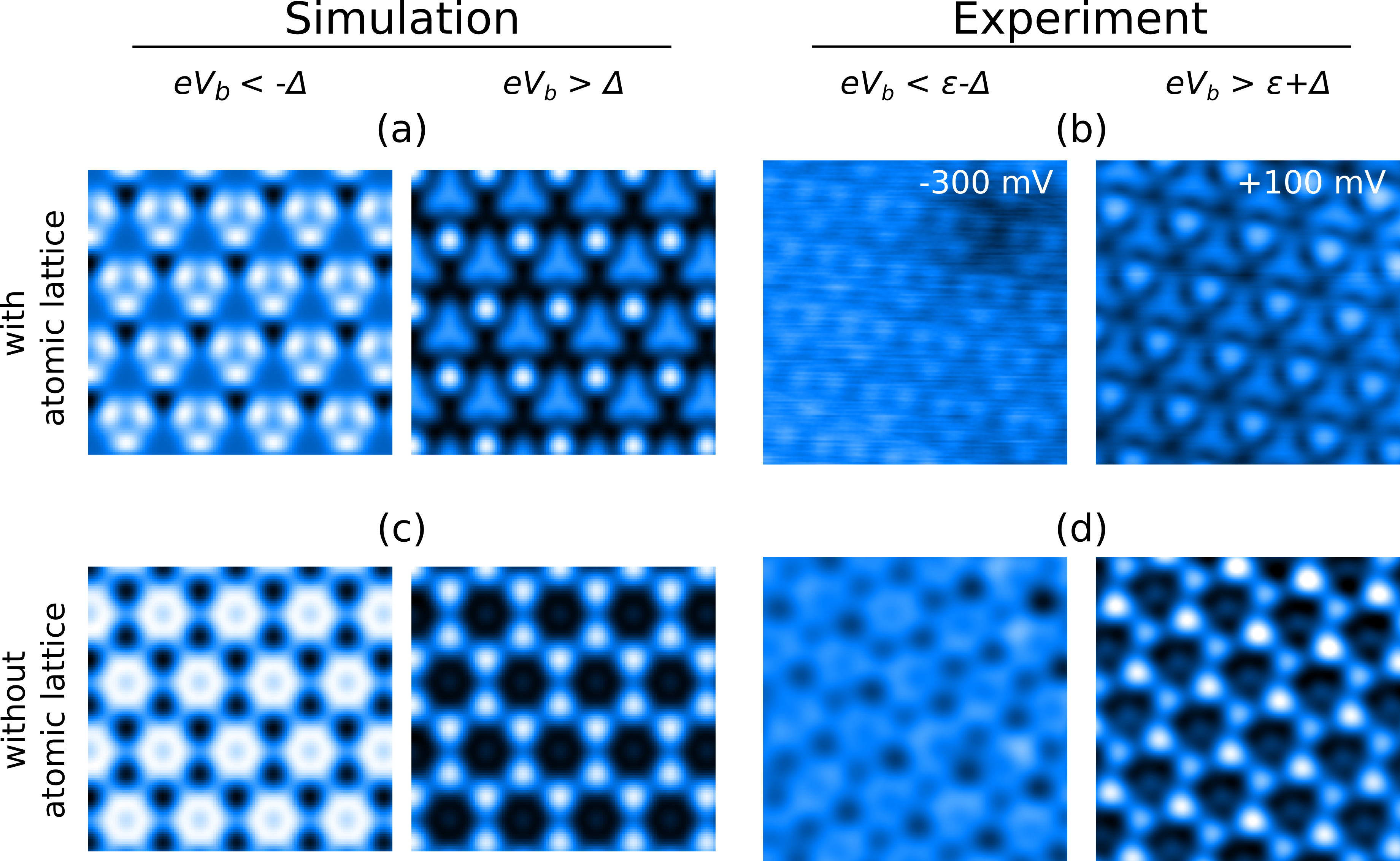}
\caption{Topographic STM patterns of a $2a \times 2a$ CDW at $V_b$ below and above the CDW gap. (a) STM topographies simulated using Eq.~\eqref{eq:sampledos2d} at $V_b = \pm 5\Delta$ for a CDW gap centered at $E_F$ ($\epsilon=0$), including the atomic lattice, and the first and second order CDW components whose phases are set to $2 \pi/3$ and $\pi$, respectively. 
The $\mathbf{k}_i$ wave vectors of the first order components are oriented along $\pi/2+(i-1)2\pi/3$ for $i=1,2,3$. (b) Experimental STM topographies measured below and above $\Delta$ on 1\textit{T}-TiSe$_2$ (adapted from Spera et al~\cite{Spera2020}). (c), (d) Data in (a) and (b), respectively, filtered to their first and second order CDW components. We refer to the pattern above (below) $E_F$ as pattern A (pattern B).}
\label{fig:simulation2d}
\end{figure}

Equations \eqref{eq:bardeensimp} and \eqref{eq:sampledos2d} enable the modeling of all the topographic and spectroscopic maps discussed here. Consistent with topographic STM images of the $2a \times 2a$ CDW in 1\textit{T}-TiSe$_2$ (Fig.~\ref{fig:simulation2d}b), our simulations reveal two characteristic CDW patterns related by contrast inversion, irrespective of the shift $\epsilon$ of the gap. Fig.~\ref{fig:simulation2d}a shows representative maps of the two patterns obtained assuming a gap centered at $E_F$ and opposite bias polarities ($V_b=\pm 5\Delta$).

\section{Results}
The CDW ground state in the classic Peierls mechanism is described as a lattice distortion and a concomitant local charge redistribution such that the reconstructed LDOS below and above the CDW gap are staggered, as illustrated in Fig.~\ref{fig:cdwdos} for a 1D CDW. 
Thus, one expects a contrast inversion between STM and STS
maps below and above $\Delta$, which has indeed been observed in
quasi-1D~\cite{Brun2009} and 2D~\cite{Carpinelli1996,Mallet1999} systems.
Proper visualization of this contrast inversion in 1\textit{T}-TiSe$_2$ requires to consider not only the first order CDW Fourier components, corresponding to the doubling of the lattice parameters, but also their differences (i.e. second order Fourier components)~\cite{Spera2020}. Therefore, we filter all STM and STS maps to the first and second order Fourier components for the analysis as shown in Figs.~\ref{fig:simulation2d}c and d. From here on, we refer to the imaging textures above and below the CDW gap as \textit{pattern A} and \textit{pattern B}, respectively. 

\begin{figure*}
\includegraphics[width=0.85\textwidth] {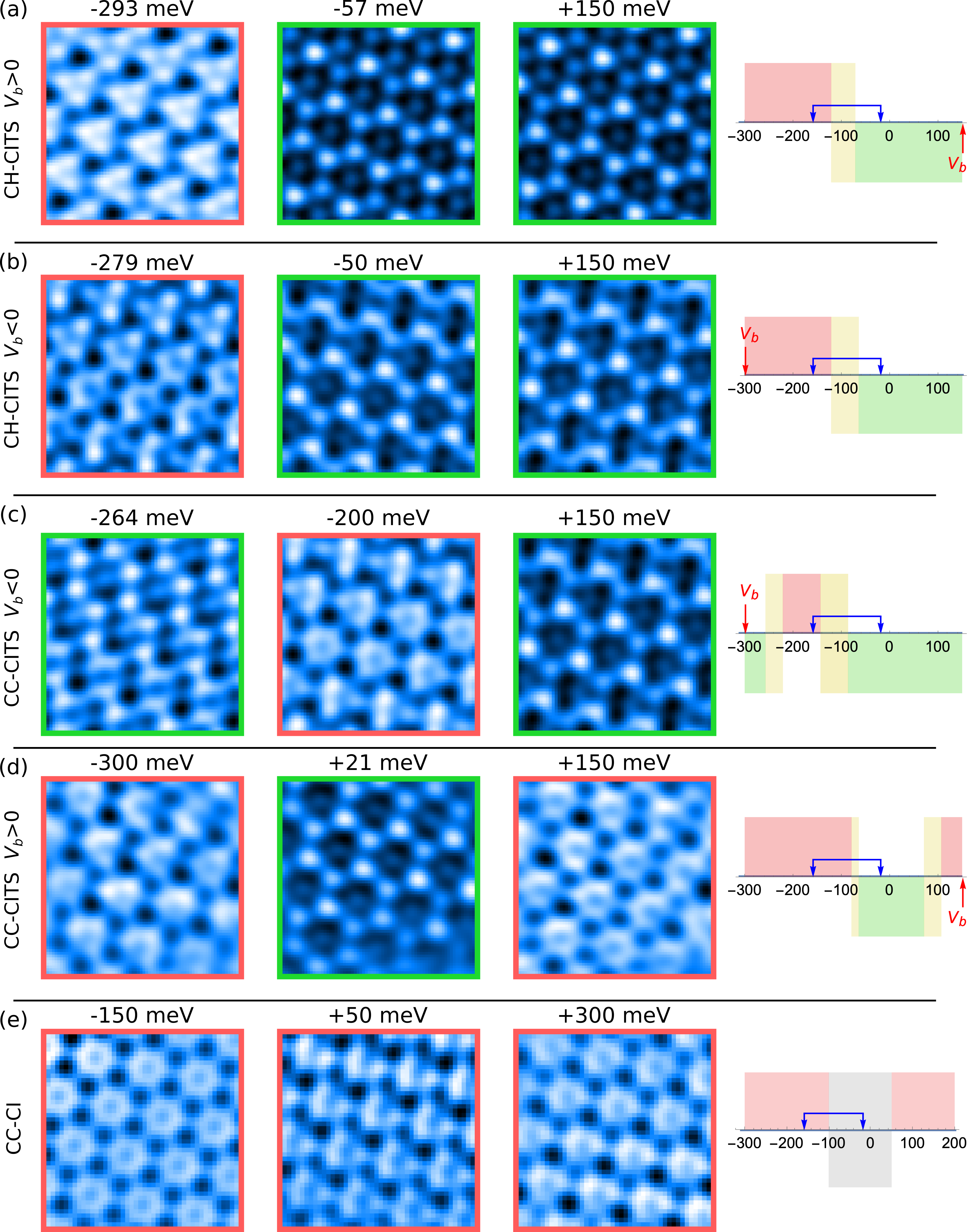}
\caption{Experimental STS data filtered to the first and second order CDW components as a function of the imaging mode: (a) CH-CITS at $V_b=-300\,\mathrm{mV}$, (b) CH-CITS at $V_b=+150\,\mathrm{mV}$, (c) CC-CITS at $V_b=-300\,\mathrm{mV}$, (d) CC-CITS at $V_b=+150\,\mathrm{mV}$ and (e) CC-CI.
The last column depicts graphically the energy ranges where pattern A (color-coded green) and pattern B (color-coded red) are observed for each imaging mode. Yellow depicts bias ranges where the CDW amplitude is too faint to be unambiguously identified, gray indicates the bias range where no data could be acquired, and the blue arrowsat $-90 \pm 70\,\mathrm{mV}$ mark the gap edges as found in Ref.~\cite{Spera2020}.
For all the CITS modes (a-d), the energy of each 2D cut from the full dataset is specified on top of the corresponding image, and the stabilization biases, marked with a red arrow, are outside the gap. For the CC-CI mode (e) the energy cut coincides with the stabilization bias.
Images size: $2.5 \times 2.5 ~\mathrm{nm}^2$.}
\label{fig:DataOverview}
\end{figure*}

Typical CDW patterns observed in conductance maps measured on 1\textit{T}-TiSe$_2$ at different energies and using different STS acquisition modes are shown in Fig.~\ref{fig:DataOverview}. They are all filtered to the first and second order CDW fourier components, and all show either pattern A or pattern B defined earlier and highlighted in green and red, respectively. A single and clearly defined CCI is resolved in the CH-CITS images in Fig.~\ref{fig:DataOverview}(a,b). Analysing the entire CH-CITS data set, we observe pattern B at all energies below $-120\,\mathrm{meV}$ (red shaded range) and pattern A at all energies above $-80\,\mathrm{meV}$ (green shaded range). The yellow-shaded energy range is where the CDW contrast is too faint to be unambiguously identified. The picture emerging from the CH-CITS data in Fig.~\ref{fig:DataOverview}(a,b) is a well resolved CDW contrast above and below a single CDW gap, which are related to each other through contrast inversion. The region in-between, where the CDW contrast is not resolved, corresponds to the CDW gap, which is centred near $-90\,\mathrm{meV}$ with an amplitude of about $70\,\mathrm{meV}$. More importantly, these observations are independent of the set-point bias polarity.

All the other STS maps in Fig.~\ref{fig:DataOverview} obtained using constant-current imaging modes reveal a very different picture. In CC-CITS, an additional CCI appears at the same polarity as the set-point bias voltage (Fig.~\ref{fig:DataOverview}(c,d)). On the other hand, no CCI is resolved in the CC-CI dataset in Fig.~\ref{fig:DataOverview}(e). It only reveals pattern B, suggesting there is no CDW gap in the displayed energy range --the gray-shaded area corresponds to a range where stable conductance imaging is impossible due to the low bias voltage. 

The conclusion at this stage is that only data obtained using CH-CITS enable a direct and unambiguous extraction of the CDW gap amplitude and position in energy (Fig.~\ref{fig:DataOverview}(a,b)) from the contrast inversion across the CDW gap as proposed by Spera et al.~\cite{Spera2020}. Data obtained using the more common constant current imaging modes must be examined much more carefully. In that case, CCIs can be misleading since they may be observed at energies unrelated to a CDW gap or may be totally absent.

\section{Discussion}
\begin{figure}
\includegraphics[width=0.9\columnwidth] {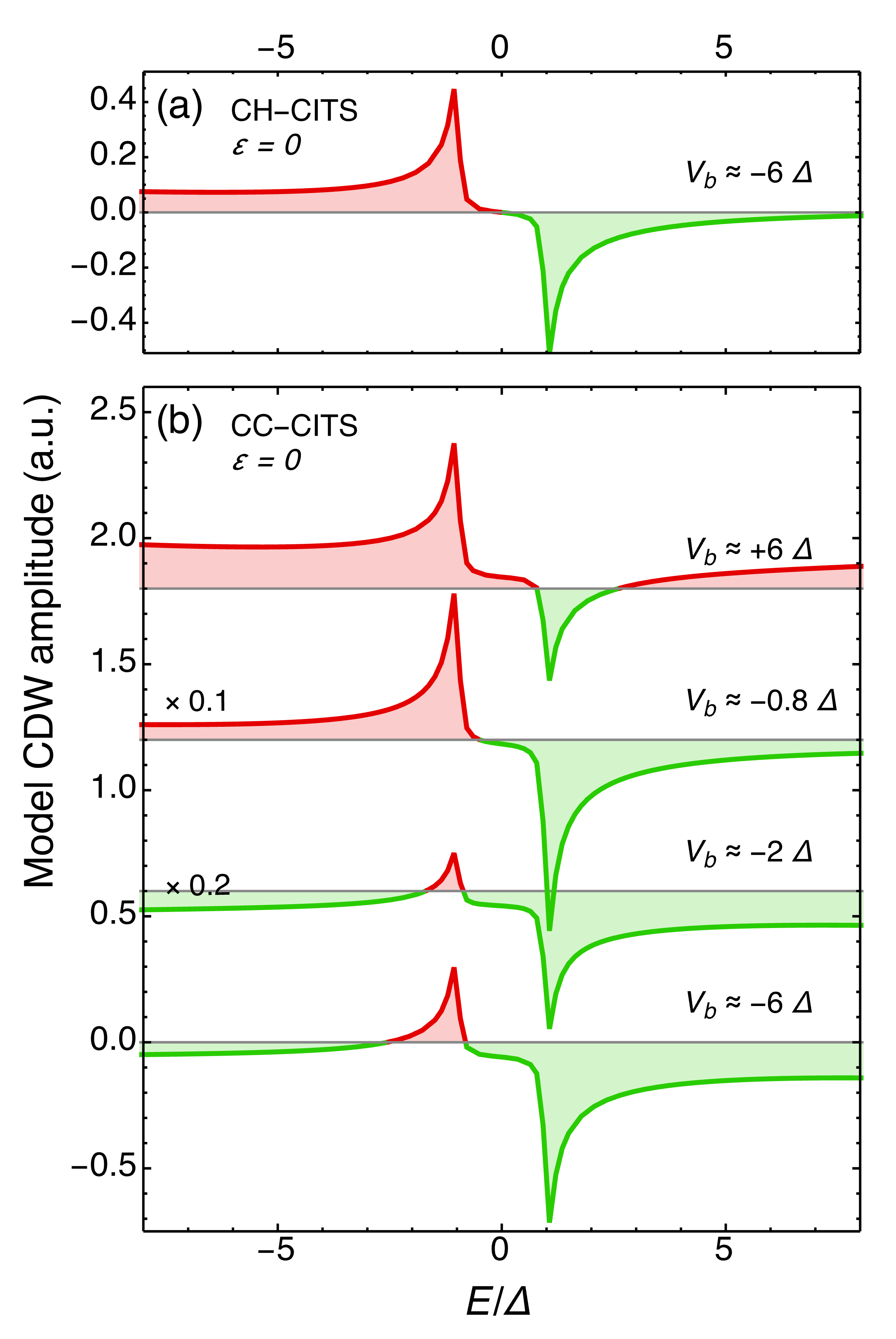}
\caption{CDW contrast amplitude calculated for different set points $V_b$ as a function of energy normalized to the CDW gap ($\Delta$) for a gap centered at $E_F$. Amplitudes expected in (a) CH-CITS and (b) CC-CITS maps, with negative (positive) values corresponding to the CDW pattern A (B) shaded in green (red). 
In the CH-CITS mode (a), the CCI does not depend on the stabilization bias, hence only one representative bias set-point is shown.
Top three curves in panel (b) are offset for clarity.}
\label{fig:citsSimNoShift}
\end{figure}

\begin{figure}
\includegraphics[width=0.9\columnwidth] {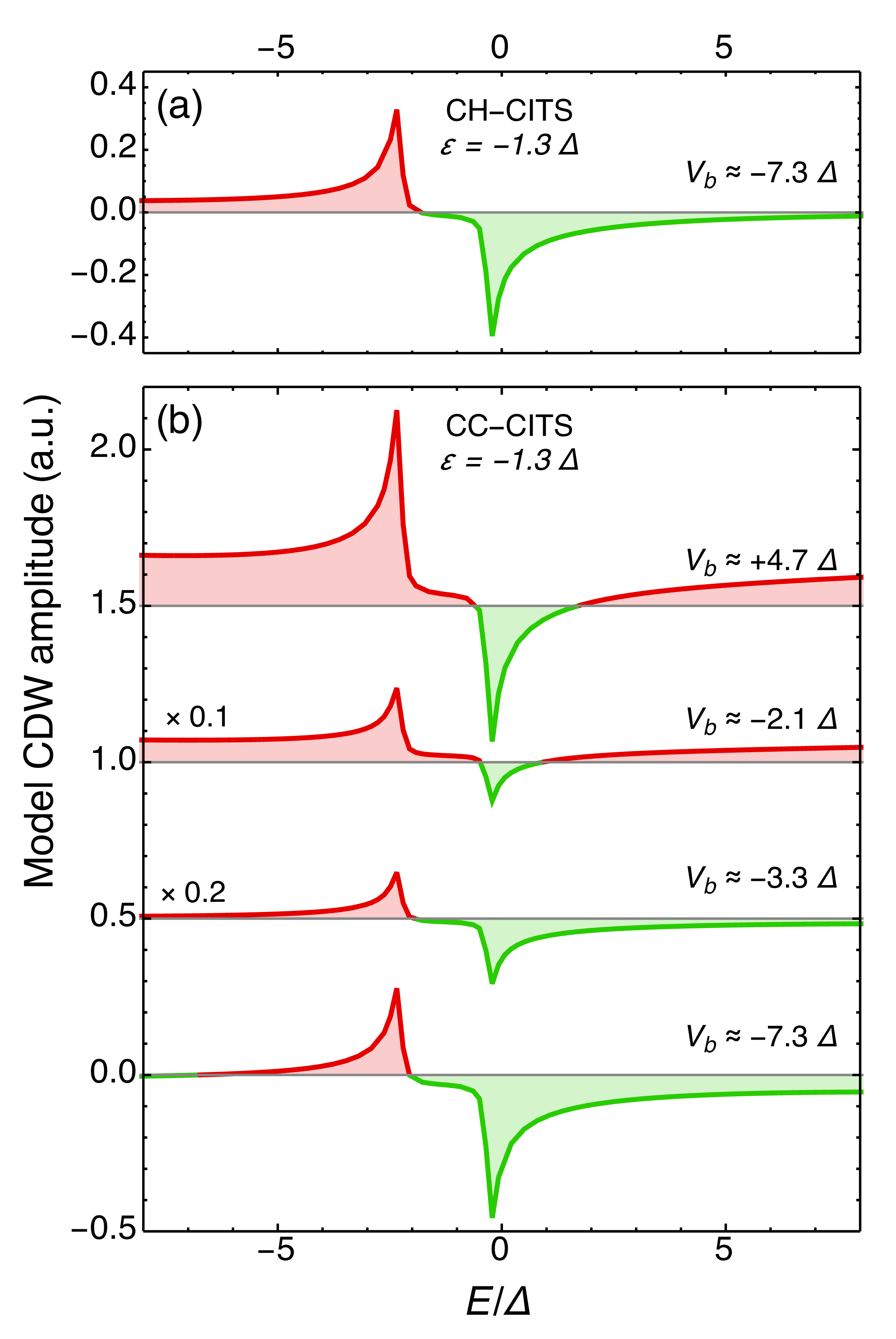}
\caption{CDW contrast amplitude calculated for different set points $V_b$ as a function of energy normalized to the CDW gap ($\Delta$) for a gap shifted by $\epsilon = -1.3 \Delta$, i.e. below $E_F$. Amplitudes expected in (a) CH-CITS and (b) CC-CITS maps, with negative (positive) values corresponding to the CDW pattern A (B) shaded in green (red). Top three curves in panel (b) are offset for clarity.
}
\label{fig:citsSim13Delta}
\end{figure}

To understand our experimental observations, we simulate the CDW contrast using the 1D model described in Section~\ref{sec:model}. Calculating the amplitude of CH-CITS maps is straightforward by evaluating Eq.~\eqref{eq:bardeensimp} for fixed tip-sample separations ($z=cte$) as a function of energy $eV$ and position $x$. Spatially resolved $dI/dV(x,V)|_{z=cte}$ grids are obtained by numerically differentiating the calculated $I(x,V)|_{z=cte}$ spectra. Computing the CC-CITS maps for a given set-point is more complex. First, we determine the topographic profile $z(x)|_{V=V_b,I=I_{set}}$ by numerically solving Eq.~\eqref{eq:bardeensimp}, which will be used to define the tip-sample separation required to calculate $I(x,V)|_{z=z(x)}$ at each position $x$. To obtain the conductance map, we determine $dI/dV(x,V)$ based on $I(x,V)$ calculated over a small energy range around $V$ at each position $x$. The result is a four-dimensional dataset from which we can extract the CDW contrast for each selected energy $eV$. The process is repeated for different sample biases $V_b$. The resulting CDW amplitudes are summarized in Fig.~\ref{fig:citsSimNoShift} for a CDW gap $\Delta$ opening at $E_F$, and in Fig.~\ref{fig:citsSim13Delta} for a CDW gap shifted below $E_F$ by $1.3\Delta$. We assign negative values and a green color (positive values and a red color) to the amplitudes corresponding to the CDW pattern A (pattern B). 

Despite its simplicity, our model agrees well with our experimental observations in Fig.~\ref{fig:DataOverview}. The CDW gap can be readily identified in CH-CITS images when it opens at $E_F$ in Fig.~\ref{fig:citsSimNoShift}(a). Here, the CDW modulation amplitude is maximal and inverted at the gap edges, with a vanishing amplitude inside the gap. Such direct correspondence between CCI and CDW gap is lost in CC-CITS images. Fig.~\ref{fig:citsSimNoShift}(b) shows that CCI no longer occurs at $\Delta$ and that there is an additional CCI above (below) $E_F$ when $V_b>0$ ($V_b<0$). The latter is unrelated to the CDW phase and its position depends on $V_b$. Consequently, any CCI taking place at the same polarity as $V_b$ needs to be considered carefully in constant current data sets as it is likely a set-point effect unrelated to the CDW ground state. According to our model, a single CCI is found in CC-CITS maps only when $|V_b|<\Delta$, although its position in energy does not match the position of the CDW gap.

The situation is very similar when the CDW gap opens away from $E_F$. To compare our model calculation with the experimental data in Fig.~\ref{fig:DataOverview}, we consider a gap shifted by $\epsilon = -1.3 \Delta$, i.e. below the fermi level~\cite{Spera2020}. CH-CITS data provide an accurate measure of $\Delta$, with a single CCI, and maximal and inverted CDW modulation amplitudes at the gap edges (Fig.\ref{fig:citsSim13Delta}a). On the other hand, CC-CITS maps again suffer from CCIs shifting with $V_b$ and an additional CCI when $V_b$ is significantly outside $\Delta$ (Fig.\ref{fig:citsSim13Delta}b).  

\begin{figure}
\includegraphics[width=0.9\columnwidth] {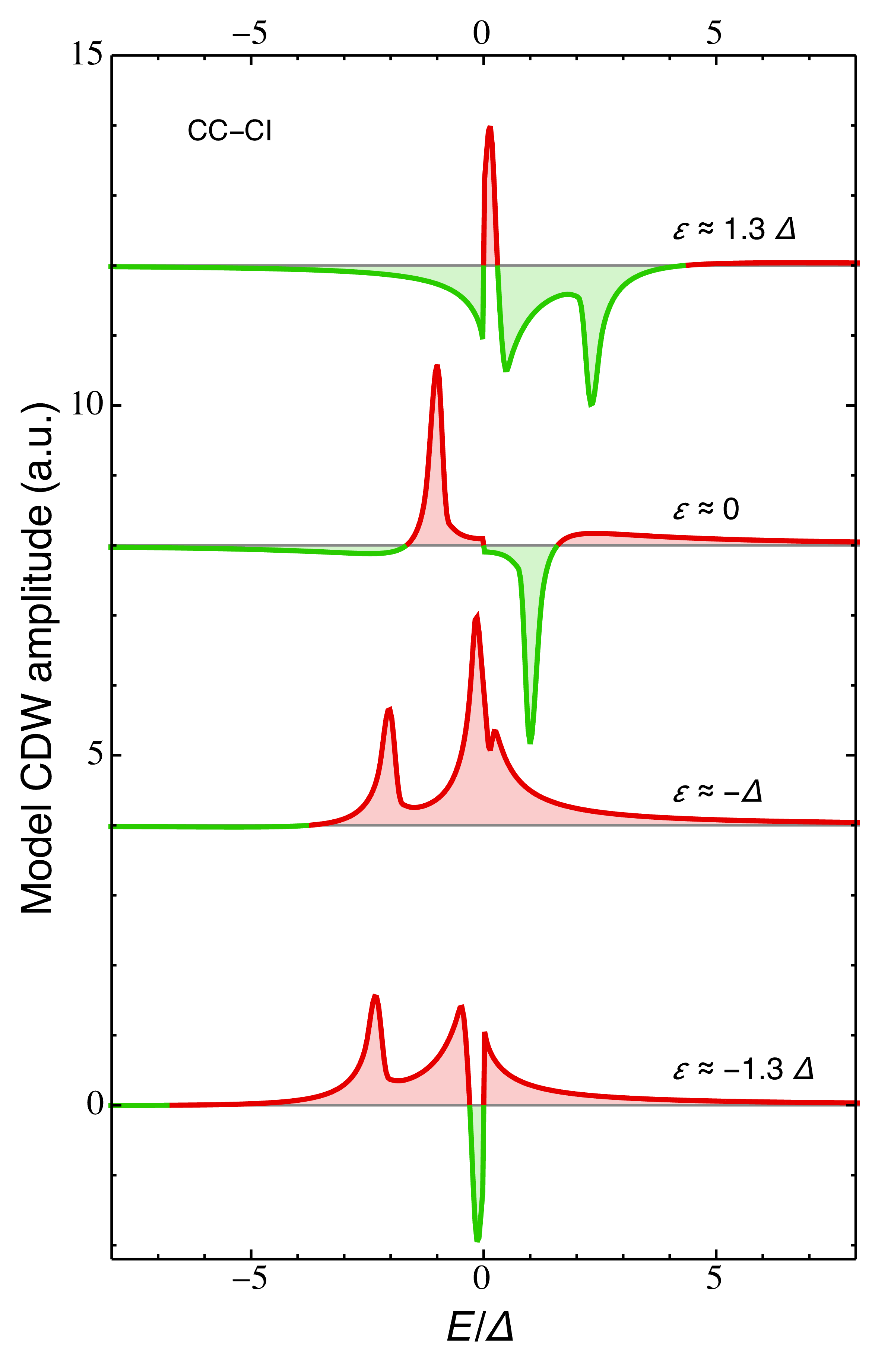}
\caption{Calculated CDW contrast amplitude expected in CC-CI maps depending on the position of the gap as a function of energy normalized to the CDW gap ($\Delta$). Negative green-shaded values correspond to the CDW pattern A, and positive red-shaded values correspond to the CDW pattern B. The curves are offset for clarity.}
\label{fig:condmapSim}
\end{figure}

Finally, our simple model also correctly reproduces the absence of CCI observed in CC-CI maps of TiSe$_2$ (Fig.~\ref{fig:DataOverview}e), where pattern B is observed at all energies for a CDW gap shifted by $\epsilon = -1.3 \Delta$ in Fig.~\ref{fig:condmapSim}. The small region near $E_F$, where pattern A is expected in this case, is not accessible to CC-CI experiments because $V_b$ is too small. Figure~\ref{fig:condmapSim} further shows the very different CC-CI imaging bias dependence of the observed CDW patterns depending on the position of the CDW gap with respect to $E_F$. When the CDW gap is shifted below $E_F$, CC-CI predominantly shows pattern B, while pattern A is dominant when the CDW gap is shifted above $E_F$. Both patterns are only resolved when $E_F$ is within the CDW gap ($\lvert \epsilon \rvert <\Delta/2$, Fig.~\ref{fig:condmapSim}), although with two additional CCIs. 

\section{Conclusion}
Measuring the CDW gap amplitude by tunneling spectroscopy is notoriously difficult and often inconclusive. Contrast inversion across the CDW gap in topographic and spectroscopic scanning tunneling images has been proposed as an alternative for identifying spectral features with the CDW gap \cite{Spera2020}. However, the correct assignment of any observed CCI to a CDW gap requires a careful analysis of the imaging mode deployed. The model analysis presented here fully agrees with the findings of Tresca et al.\cite{Tresca2023}, in particular that CC-CI is inappropriate to identify CDW patterns. Our dataset and model calculations further suggest that some of the spectral features identified as CDW gaps based on CCI in constant current conductance maps in bulk 1\textit{T}-TaS$_2$~\cite{Cho2015}, in monolayer VS$_2$~~\cite{vanEfferen2021}, in Cr(001)~\cite{Hu2022}, or in a metallic Kagome system~\cite{Cheng2024}, to name a few examples, may be set-point related artifacts and deserve further investigations in light of the present insight. As we demonstrate here, only constant height spectroscopic imaging provides a systematic and direct access to the CDW gap based on charge density wave contrast inversion as a function of imaging bias across the CDW gap.  

\section{Acknowledgements}
We thank A. Guipet and G. Manfrini for their technical assistance with the scanning probe instruments. This work was supported by the Swiss National Science Foundation (Division II Grant No. 182652).

\end{document}